\documentclass{elsart}

\def\axp{1RXS~J170849-400910}

\def\chan{$Chandra$}
\def\ro{$ROSAT$}
\def\asca{$ASCA$}

\usepackage{natbib}
\usepackage{graphicx}
\usepackage{amssymb}

\begin{document}
\begin{frontmatter}
\title{A Gemini Observation of the Anomalous X-ray Pulsar \axp}
\author{Samar Safi-Harb}
\author{and Jennifer West}
\address{Department of Physics and Astronomy, University of Manitoba, Winnipeg, Manitoba, R3T 2N2, CANADA
samar@physics.umanitoba.ca, westjl@cc.umanitoba.ca}

\begin{abstract}

The anomalous X-ray pulsars (AXPs) represent a growing class of neutron stars discovered at X-ray energies. While the nature of their multi-wavelength
emission mechanism is still under debate,  evidence has been recently accumulating  in favor of their magnetar nature.
Their study in the optical and infrared (IR) wavelengths has recently opened a new window to constrain the proposed models.
We here present a brief overview of AXPs and
our Gemini-South observation of \axp, which is 
a relatively bright AXP  discovered with \ro\ and later found to be an 11 s X-ray pulsar by \asca. 
The observation was taken with the near-IR imager Flamingos in J (1.25 $\mu$m), H (1.65 $\mu$m), and K$_s$ (2.15 $\mu$m). We confirm the recent detection by \citet{israel2003} of a source coincident with the \chan\ source (candidate `A'). Our derived magnitudes of J = 20.6 (0.2), H = 18.6 (0.2), and
Ks = 17.1 (0.2) are consistent with those derived by \citet{israel2003}, and indicate that if this source is indeed the IR counterpart to \axp, then there is no evidence of variability from this AXP. 
However, given the lack of IR variability and the relatively high IR to X-ray flux of this source when compared to the other AXPs, we conclude that this source is unlikely
the counterpart of the AXP, and that the other source (candidate `B') within the \chan\ error circle should not be ruled out as the counterpart. 
Further monitoring of these sources and a 
deep observation 
of this complex field are needed to confirm the nature of these sources and their
association with the AXP.
\end{abstract}

\end{frontmatter}

\section{Introduction}
\label{section:intro}
The `Anomalous X-ray Pulsars' (AXPs) represent a growing class of pulsars 
whose X-ray luminosities ($\sim$10$^{34}$-10$^{36}$ erg~s$^{-1}$) can not be accounted for by their spin-down power nor by conventional binary accretion models, thus the name `anomalous'. Currently, we know of 8 AXPs, one of which is a transient AXP (XTE~1810-197) and two are AXP candidates
(CXOU~J0110043.1-721134 and AX~1845-0258).
Typified by 1E2259+586, the first AXP discovered in the SNR CTB~109~\citep{gregory1980},
AXPs share the following properties:
\begin{enumerate}
\item{they are slow X-ray rotators (when compared to the rotation-powered Crab-like pulsars) with periods constrained to the narrow range of 5--12 seconds,
and unlike the accretion-powered pulsars, they spin down;}
\item{their large spin-down rates and the association of three of them
with SNRs indicate a young age
(10$^3$--10$^5$ years, see Table~1);}
\item{they lack a detectable companion star and have a small scale height
above the Galactic Plane;}
\item{they have soft X-ray spectra (compared to the Crab-like and
accretion-powered pulsars) characterized by a two-component blackbody+power law;}
\item{they do not have radio counterparts; but  half the sample has faint optical and/or IR counterparts (see Table~1).}
\end{enumerate}

The two competing models proposed to describe the emission mechanism from these sources have been the accretion and `magnetar' models.

In the accretion model, typical high-mass X-ray binaries are ruled out based on the
AXPs' relatively low X-ray luminosities,  soft X-ray spectra,  lack of Doppler shifts associated with binary orbital motion, and the absence of bright optical counterparts.
However, models which involve accretion from a
fossil disk, established from matter falling back onto the neutron star following its birth, have not been ruled out
and have been favored by some authors~\citep{chatterjee2000, alpar2001, marsden2001}.
These accretion models are particularly successful in explaining the period clustering of AXPs and their  persistent X-ray emission.
 However, they are problematic in explaining the bursts energetics and
  the optical pulsations from 4U~0142+614 \citep{kern2002};
  see however \citet{ertan2004} who argue that the observed optical pulsed emission from
  4U~0142+614 can be explained within the disk model.
    
Evidence has been recently accumulating in favor of the `magnetar' model, which was proposed  to explain the properties of their relatives, the
Soft Gamma-Ray Repeaters (SGRs).
In this model~\citep{duncan1992, thompson1995, colpi2000},
AXPs and SGRs 
have inferred magnetic fields of $\sim$10$^{14-15}$ G, at least two orders of magnitude higher than the Crab
and an order of magnitude larger than the quantum critical field.  Their large spin-down torques are provided by the magnetic dipole radiation. The 
super-Eddington bursts result from the release of magnetic energy through instabilities from inside the neutron star. The persistent thermal X-rays were suggested to be due to the surface heating by magnetic field decay, while the nonthermal X-rays could be due to acceleration of particles by the Alfv\'en waves on the neuron star surface.
The magnetar model has been successful in explaining most of the AXPs properties,
including  the period clustering of these sources when the magnetic field decays significantly on a timescale of $\sim$10$^{4}$~years~\citep{colpi2000}.
However, the emission at optical and IR wavelengths is yet to be further explored in the light of the recent observations (see \S2).

\section{Optical and IR Observations:}

As of today, 
4(5) AXPs have been recently detected in the optical or IR.
In Table~1, we summarize their properties including their spin periods, their
dipole magnetic fields ($B$$\sim$3.2$\times$10$^{19}$($P\dot{P})^{0.5}$~Gauss), their spin down ages ($\tau$=$P/2\dot{P}$), their association with SNRs, and the 
magnitudes of their proposed counterparts.
Optical and IR observations have been recently driving 
a wealth of theoretical modeling in order to understand their emission
mechanisms at those wavelengths.

In the accretion model, energy is dissipated by viscous processes in the accretion disks. \citet{perna2000} computed the theoretical predictions for the optical and IR emission from the accretion model established by fall-back following a supernova explosion \citep{chatterjee2000}. 
While the predicted magnitudes are brighter than the observed values, \citet{perna2000} noted that the discrepancy could be explained by the sensitivity of the predicted fluxes (especially in the optical regime) to the value of the inner part of the disk, and that the emission at the near-IR wavelengths should originate from the outermost regions of the disk. \citet{perna2000} also argued that the spin-down rate from accretion models is identical to that predicted by the magnetar model. Therefore, the ability to distinguish between the two models lies in searching for disk emission at wavelengths longer than X-rays. 
\citet{mereghetti2002} noted that the geometry and size of the disks are too uncertain to definitely rule out the accretion model. 

In the magnetar model, there hasn't been much theoretical prediction of the optical/IR emission; and it is only recently that \"Ozel (2004) argued that, in the magnetar model,  the IR emission can not be due to thermal emission from the neutron star surface, but that it is synchrotron emission from the magnetosphere. In this model,
 the IR to X-ray flux is expected to be correlated with the spin-down energy of the AXP. However,
\citet{rea2004} found that the IR emission from XTE~1810-197 does not follow this general trend.

It is likely that `hybrid' models which involve a neutron star with a magnetar field strength in the higher multipoles, and a dipole field in the 10$^{12}$~G interacting with a fallback disk, are more realistic in explaining all AXP properties (Eksi \& Alpar 2003). The higher multipole magnetic field would explain the bursting properties of the AXPs, and the AXP period clustering and X-ray/IR correlation can be explained in the accretion model.
This `hybrid' model has been recently supported by the IR observations of XTE~1810-197 \citep{rea2004}.

\vspace{-0.3cm}

\section{\axp}

\subsection{Previous observations}
\axp\ was first discovered with the \ro\ X-ray satellite \citep{voges1996}, and later found to be an 11~s pulsar with \asca\ \citep{sugizaki1997}.
This AXP is one of the brightest AXPs with an X-ray luminosity of
$\sim$2$\times$10$^{35}$~erg~s$^{-1}$ at a distance of 5~kpc.

This source lies in a complex region of the sky and in the neighborhood of the supernova remnant G346.5--0.1. The association between the AXP and the SNR
is however unlikely \citep{gaensler2001}.
No radio counterpart to \axp\ has been found and a 5$\sigma$ upper limit of 3~mJy was
placed~\citep{israel2003, gaensler2001}.
Recent \chan\ observations have located the source at
$\alpha$=17$^h$ 08$^m$ 46$^s$.87, $\delta$=-40$^o$ 08$^{\prime}$ 52$^{\prime\prime}$.44 (J2000) 
with an uncertainty circle radius of 0$^{\prime\prime}$.8
(90\% confidence level) consistent with the 
\ro\ HRI observation \citep{israel2003}.

BeppoSAX observations of \axp\ reported on the first detection of an absorption
line near 8.1 keV, which if interpreted as a proton cyclotron line, would imply
a magnetic field value of B=1.6$\times$10$^{15}$~G.  This value is higher but
not inconsistent with the magnetic field derived from the spin-down of the source
(B$\sim$4.6$\times$10$^{14}$ Gauss), a magnetar-strength value.
While no bursting activity has been yet reported from this AXP,
it was observed to glitch~\citep{dallosso2003, kaspi2003a}.

Recently, optical and IR observations were performed with
the 3.6~m ESO telescope in La Sille, Chile. Additional IR data were obtained
with the 3.6~m CFHT (Mauna Kea, Hawai).
While there was no optical detection within the \chan\ circle down to a limiting
magnitude in R of 26.5, two faint objects were detected in the IR \citep{israel2003}.
The two sources, named A \& B, have the following magnitudes:
K$^{\prime}$=17.53$\pm$0.02, H=18.85$\pm$0.05, K$_s$=17.3$\pm$0.1\footnote{K$_s$ was reported as 18.3 in \citet{israel2003}, however the magnitudes in K and K$_s$ should be similar and so we corrected for the typo (Israel, private communication).}  for  object A; and
K$^{\prime}$=20.0$\pm$0.08, H=20.43$\pm$0.07 for object B.
\citet{israel2003} argue that object A is most likely the counterpart of \axp.

\subsection{The Gemini observation:}
This Gemini observation was part of an approved Gemini program to search for and study the IR counterparts of AXPs.  Due to unfortunate problems with the Near Infrared Imager (NIRI)  in both cycles 2001B and 2002B, the Gemini-North observations of AXPs 1E~2259+586, 1E~1841-045,  and AX~J1845-0258 did not take place.  

\axp\ was however observed using the Gemini-South FLAMINGOS I near-IR (1--2.5~$\mu$m) multi-object spectrograph and imager, built at the University of Florida. On
2002, July 11-12, 40 individual 45 s images in the K$_s$ (1.99 -- 2.30 $\mu$m) 
filter were obtained for a total exposure time of 1,800 s. On 2002, July 12-13,
37 individual 60~s images in the H (1.49--1.78 $\mu$m) filter for a total exposure time of 2,220~s and 25 individual 90~s images in the J (1.15--1.33 $\mu$m) filter for a total exposure time of 2,250~s were obtained. July 12-13 had photometric weather. The images have a scale of 0.078 arcsec pixels with a field of view of 2.6$^{\prime}$$\times$2.6$^{\prime}$.

The individual images were sky-subtracted, flat-fielded and then combined using the Gemini IRAF package following the standard reduction techniques for FLAMINGOS data. The FWHM of the combined observations is 0.5$^{\prime\prime}$--0.6$^{\prime\prime}$ for both nights.

The Persson IR standard stars \citep{persson1998} S754 and S294 were observed on July 2002, 11-12, and S279 was observed on 2002, July 12-13.
These data were reduced as described above and used for calibration. Photometry were obtained using DAOPHOT II \citep{stetson1987}. The limiting magnitude was found to be 19.3 (K$_s$), 21.3 (H), and 22.0 (J).

\section{Results and Discussion:}
As shown in Fig.~1, we confirm \citet{israel2003} detection of candidate
`A',  an IR source located within the 0.8$^{\prime\prime}$ uncertainty region of the position of \axp, as determined by the \chan\ X-ray observation.  

Table 1 displays the magnitudes and colour indices for candidate `A' and compares them with those found by \citet{israel2003}. It should be noted that the resolution of the Gemini data is such that candidates `A' and `B' are not resolved and therefore the colour indices determined for candidate `A' could be contaminated slightly, but not significantly, by candidate `B'.
Using CFHT, Israel et al. (2003) measured the H magnitude of both candidates `A' and `B' to be 18.85 +/- 0.05 and 20.43 +/- 0.07, respectively. The combined magnitude given these two measurements would be 18.6 and therefore  we estimate the contamination in magnitude to be ~0.2-0.3.
This is consistent with the difference in magnitudes between the
Israel et al. (2003) measurement and ours (Table~2).

While IR variability seems to be a common property of AXPs,
\citet{israel2003} reported no variability in \axp\ for  the observations taken 1999 September 15-16, 2001 May 26, \& 2002 February 19.
As well, our Gemini magnitudes agree with those of \citet{israel2003} in the J, H, and Ks bands, within error (Table~2). We conclude that there is also no  evidence of variability  between the Israel et al. observations and ours.

While the absence of IR variability could be explained by the (so far) no detection of bursts from this AXP,
we believe that a more likely interpretation of
the absence of IR variability over $\sim$3~years
 is that candidate `A'  is not the true counterpart to the AXP.
Unless \axp\ is an anomalous AXP, this conclusion is further supported by the very small X-ray to IR flux ratio ($\sim$500) compared to that observed for all other AXPs ($>$1000).  We therefore conclude that
  candidate `B' or other fainter (undetected) stars in this crowded field should not be ruled out as a potential counterpart to \axp. Future deep observations and simultaneous monitoring in the IR and X-rays  are needed to confirm our conclusions.

\vspace{0.5cm}
SS-H is an NSERC University Faculty Award fellow and acknowledges support by an NSERC Discovery grant and a University of Manitoba Research Grant.\\
This research is based on observations obtained at the Gemini Observatory, which is operated by the Association of Universities for Research in Astronomy, Inc., under a cooperative agreement with the NSF on behalf of the Gemini partnership: the National Science Foundation (United States), the Particle Physics and Astronomy Research Council (United Kingdom), the National Research Council (Canada), CONICYT (Chile), the Australian Research Council (Australia),


\begin{table*}[htbp]
\begin{center}
\begin{tabular}{cccccccc} \hline
\footnotesize
AXP & P & B  & $\tau$ & SNR? & optical/IR counterpart$^{[*]}$ \\
      & (s) & (10$^{14}$~G) &  (10$^3$ yrs)  & & & &  \\ \hline
XTE1810-197 & 5.54 &  2.6 & 7.6 & -- & I$>$24.3,
H=21.3--22.0\\
 & & & & &  K$_s$=20.3--20.8$^{[1]}$\\
1E~1048.1-5937 & 6.45 &  5.0 & 2.7 & -- & I=26.6, J=21.7, H=20.8\\
 & & & & & K$_s$=19.4--21.2$^{[2]}$\\
AX~1845-0258 & 6.97 & ? & ? & G29.6+0.1 & --  \\
1E~2259+586 & 6.98 & 0.59 & 230 & CTB~109 & J, I, R$>$23.8, 25.6, 26.4 \\
 & & & & & K$_s$=20.4--21.7$^{[3]}$\\
CXOU~0110043.1-721134 & 8.0 & ? & ? & -- &  -- \\
4U~0142+615 & 8.69 & 1.3 & 72 & -- & R=24.98$^{[4]}$ \\
1RXS~J170849-400910 & 11.0 & 4.6 & 9.4 & -- &  (see Table~2)$^{[5]}$\\
1E~1841-045 & 11.8 & 7.0 & 4.6 & Kes~73 & -- \\
\hline
\end{tabular}
\end{center}
[*] The references that follow correspond to the optical and infrared
observations [1] Israel et al. (2004); Rea et al. (2004) [2] Wang \& Chakrabarty (2002);  Durant, van Kerkwijk \& Hulleman (2003) [3]  Hulleman et al. (2001); Kaspi et al. (2003);
[4] Hulleman et al. (2000); [5] Israel et al. (2003) and this work.\\
\caption{Summary of AXPs and their proposed optical/IR counterparts.}
\label{table:axps}
\end{table*}

\clearpage

\begin{table*}[htbp]
\begin{center}
\begin{tabular}{ccc} \hline
 magnitude/reference     & this work & Israel et al. (2003) \\
 \hline
 J & 20.6$\pm$0.2 & 20.9$\pm$0.1\\
 H & 18.6$\pm$0.2 & 18.85$\pm$0.05\\
 --	& & 18.6$\pm$0.1 \\
K$_s$ & 17.1$\pm$0.2 & 17.3$\pm$0.1$^{(a)}$\\
K$^{\prime}$ & & 17.53$\pm$0.02\\
J-H & 2.0$\pm$0.4 & 2.3$\pm$0.2\\
J-K$_s$$^{(a)}$ & 3.5$\pm$0.4 & 3.6$\pm$0.2 \\
J-K$^{\prime}$ & & 3.4$\pm$0.1 \\
H-K$_s$$^{(a)}$ & 1.5$\pm$0.4 & 1.6$\pm$0.2\\
	-- & & 1.3$\pm$0.2 \\
H-K$^{\prime}$ &  & 1.32$\pm$0.07 \\
-- &  & 1.1$\pm$0.1 \\
\hline
\end{tabular}

$^{(a)}$ after correcting for the typo in Israel et al. (2003); see text
for details.
\end{center}
\caption{Infrared magnitudes of candidate `A'.}
\label{table:magnitudes}
\end{table*}

\clearpage

\begin{figure*}[htbp]
\begin{center}
\caption{(Left): Colour image composed of J, H, and Ks filter images coloured as shown at right. Circles shown represent 90\% error circles from ROSAT (9$^{\prime\prime}$-radius) and \chan\ (0.8$^{\prime\prime}$-radius)
as in the figure from \citet{israel2003}.
(Right): Individual filter images. White arrow points to the position of AXP Candidate `A' (Israel et al. 2003).}
\label{figure:gemini}
\end{center}
\end{figure*}

\end{document}